\documentstyle[preprint,tighten,aps]{revtex}
\begin{document}

%this is a submission to the archive astro-ph, in RevTeX, from:
%Professor R. G. Moorhouse
%Dept. of Physics and Astronomy
%Glasgow University
%GLASGOW G12 8QQ
%Scotland, U. K. 

\draft
\title{Post-Inflationary Reheating and Perturbations of the Cosmic Microwave 
Background.\footnote{Research partially funded by 
 NATO Collaborative Research Grant CRG no. 920129}}
\author{R. G. Moorhouse}
\address{Department of Physics and Astronomy,University of Glasgow,
Glasgow G12 8QQ, U.K.}
\author{A. B. Henriques, L. E. Mendes}
\address{Departamento de Fisica,Instituto Superior Tecnico,
1096 Lisbon, Portugal}
\date{\today}
\maketitle
\begin{abstract}
     We formulate a gradual dynamical transition from a power-law 
inflation era with a scalar field to a radiation era with no scalar 
field including inhomogeneous perturbations to the Friedmann- 
Robertson-Walker universe. We show that for the cosmic microwave  
background radiation fluctuations this is excellently approximated 
by a sudden transition, with application of the Lichnerowicz 
conditions, both for density and gravitational wave perturbations.  
\end{abstract} 
\narrowtext
\section{INTRODUCTION}\label{INTRO}
Post-inflationary reheating in which the energy of a scalar field of  
the inflationary era is converted into the relativistic particles 
of the radiation era is of intrinsic interest, and there may prove to 
be tests of the mechanisms of inflation and reheating. A common   
scenario is that there is a period of inflation followed by reheating,
consequently followed by the radiation era; density and graviton 
perturbations are supposed to be of quantum origin arising in the  
inflaton field(s) early in the inflationary era. It is an interesting 
question how the reheating affects the perturbations of various 
wavelengths, particularly the density perturbations. For those 
wavelengths $\lambda$ observable in the variations of the CMBR the 
phase $\Delta t/\lambda$, where $\Delta t$ is the reheat time, is  
extremely small. So the usual assumption is that for these wavelengths
the effect of the transition on the perturbations can be well 
calculated by the sudden (instantaneous) approximation. In recent  
years there has been some discussion on this point 
\cite{GRI,DER,CAL,MS,GOTZ}.It is the main purpose of this paper to 
support the usual assumption by illustration using a well known 
inflationary scenario followed by a particular reheating mechanism.

The mechanism that we use is an old one \cite{ALB}, that of 
a friction term in the scalar field equation with the corresponding 
balancing term in the radiation fluid energy equation.  We do not 
postulate this as a close representation of the actual physical 
process of reheating or defrosting, which is anyway not yet 
determined, but use it as a stand-in providing a balanced transfer  
of energy-momentum from one phase to another in a mixed phase state. 
In this latter respect it is more physically motivated than reheating 
representations which just interpolate the scale factor of the FRW 
universe \cite{GRI,CAL,MS}. 

The new feature is that the calculation of the transition from a 
universe of all scalar field to one of all radiation fluid is done 
explicitly for the perturbed as well as for the unperturbed case. In 
the former case the synchronous gauge is used for calculation. We 
find that the numerical results for the resulting perturbations in 
the radiation era are the same, to many significant figures, as those 
got in the sudden approximation by application of the Lichnerowicz 
conditions \cite{LIC} on a hypersurface of constant energy density, 
as formulated by Deruelle and Mukhanov \cite{DER}. This result serves 
to confirm the hypothesis (and its implementation) that the sudden 
phase transition happens on a hypersurface of constant energy density    
even where it is between entirely different types of physical states 
-- in this case a scalar field and a standard type of perfect fluid.

In the following section we outline the model, setting out the scalar 
potential. This is adjusted to give a period of power-law inflation, 
followed by a transition to zero scalar field, with the transition 
being mediated by a friction term. In section \ref{REHEAT} we first 
derive the resulting equations for the reheating transition in 
which the scalar field matter of the universe is wholly converted 
into a radiation fluid and the universe enters the radiation era; 
secondly we derive the equations for the development of the density 
perturbations, arisen by earlier quantum fluctuations, through the  
gradual reheating transition. Section \ref{TRANSITIONS} is mainly 
on the numerical solution of these equations to give the resulting 
radiation era parameters; for the perturbations we treat only 
wavelengths important for the observed fluctations of the cosmic
microwave background. We also derive the corresponding radiation 
era perturbation for the sudden transition approximation, so as to 
compare this with the result of the gradual transition. We show 
that the two results are the same, to extremely good accuracy. The  
computations are done in the synchronous gauge, but the comparison  
of results for the sudden and gradual transitions are done using  
gauge invariant variables \cite{MUK}. We also include a discussion of 
the effect of our gradual transition mechanism on non-physical 
so called 'synchronous gauge modes' - eliminable by a coordinate  
transformation using the residual gauge freedom of synchronous 
gauges \cite{GRI,MUK}.
       
The  metric tensor perturbations - gravitational waves - are a 
separate  and much easier case and in \ref{TRANSC} we show 
analytically that the sudden transition approximation is valid also 
in this case.

\section{MODEL FOR INFLATION AND REHEATING}\label{MODEL}
We choose a potential for the scalar field $\phi$ which gives constant
power law inflation for large $\phi$ and which, in conjunction with 
a friction term, converts the scalar field energy into a perfect 
fluid as $\phi$ decreases. We take the fluid to be the usual one of 
the radiation era, so that the pressure and density are related by 
$p=\frac{1}{3}\rho$. The potential is given by 
\begin{equation}
V = U\exp(\lambda\phi), \lbrace\lambda > 0\rbrace;\phi > \phi_{A},
\label{POTENL1} 
\end{equation}
\begin{equation}
V = V_{0}[\exp(-2\nu\phi)-2\exp(-\nu\phi)+1]; \phi < \phi_{A}. 
\label{POTENL2} 
\end{equation}
and is illustrated in the Figure for the particular case where 
$V_{0} = 1$,$\nu = \sqrt{2}$ and $\lambda = .45$.

The constants $U,V_{0},\lambda,\nu$ are adjusted so that $V,dV/d\phi$  
are continuous at $\phi=\phi_{A}$. The curvature, $d^{2}V/d\phi^{2}$, 
changes sign at $\phi_{B} < \phi_{A}$ so that it is negative for 
$\phi < \phi_{B}$ where a friction term $f\sqrt{d^{2}V/d\phi^{2}}$, 
where $f$ is an adjustable constant, is added to the equations as 
detailed below. 

The scalar field decreases with time and in the power law inflation 
region,$\phi > \phi_{A}$, the cosmic scale factor 
\begin{equation}\label{aPOWLAW}
a(\tau) \propto (\tau_{p}-\tau)^{p} \propto t^{q} 
\end{equation}
where $\tau$ is conformal and $t$ cosmic time. $\tau_{p}$ is a 
constant, $\lambda < \sqrt{2}$ and 
\begin{equation}\label{p,qPOWLAW}
p = 2/(\lambda^{2}-2) , q = 2/\lambda^{2}. 
\end{equation}
For $\phi < \phi_{A}$ the field accelerates until some 
$\phi < \phi_{B}$ where the friction becomes large enough to slow it 
down; finally the field energy is all converted into fluid energy. 

This model is just one realization of inflation followed by reheating 
and is designed to track coupled metric,scalar field and density 
perturbations through inflation to the end of reheating. These 
perturbations are supposed to arise from quantum fluctuations early 
in the inflationary era. They are accompanied by quantum fluctuations 
with graviton production giving rise to gravity waves which we shall 
likewise track through to the radiation era.

\section{INFLATION AND REHEATING EQUATIONS}\label{REHEAT} 
\subsection{The unperturbed equations}\label{REHEATA}
We write the equations for the case where the radiation fluid and the 
scalar field are coexisting; the pure scalar field era and the 
radiation era are just special cases of these equations. For 
$\phi > \phi_{A}$ the model has an analytic solution but otherwise 
a numerically computed solution is necessary, even for this 
unperturbed case.

The Einstein equations are
\begin{equation}\label{EINSTEIN} 
R_{\mu\nu} - \frac{1}{2}g_{\mu\nu} R = 
\kappa T^{s}_{\mu\nu}+\kappa T^{h}_{\mu\nu},
\end{equation}
\begin{equation} 
\kappa = 8\pi G
\end{equation}

the right hand side being the sum of the energy-momentum tensors of 
the scalar field and of the fluid,with density $\rho$ and pressure 
$p=b\rho$ and 4-velocity $u^{\mu}$:
\begin{equation}\label{Ts} 
T^{s}_{\mu\nu} = \phi_{,\mu}\phi_{,\nu}-
g_{\mu\nu}[\frac{1}{2}g^{\alpha\beta}\phi_{,\alpha}\phi_{,\beta}+V(\phi)]
\end{equation}
\begin{equation}\label{Th} 
T^{h}_{\mu\nu} = pg_{\mu\nu} + (\rho + p)u_{\mu}u_{\nu}
\end{equation}
To solve these equations the scalar field equation is also requisite.  
To the usual scalar equation we add a friction term so that it becomes
\begin{equation}\label{SCALAREQ} 
d^{2}\phi/d\tau^{2}+2aHd\phi/d\tau+a^{2}V_{1}+aV_{f}d\phi/d\tau = 0
\end{equation}
where Hubble, $H =(da/d\tau)/a^{2}$ and we have used the notations
\begin{equation}
V_{1}\equiv dV/d\phi,
\end{equation}
\begin{equation}
V_{f} = f\sqrt{d^{2}V(\phi)/d\phi^{2}},\phi < \phi_{B},
V_{f}=0,\phi > \phi_{B}.
\end{equation}

From the Einstein equations we obtain
\begin{equation}\label{HUBBLE} 
H^{2}=\frac{\kappa}{3}[(\frac{d\phi}{d\tau})^{2}/2a^{2} + V + \rho].
\end{equation}
The Bianchi identities from the Einstein equations combined with the 
scalar equation yield the fluid energy equation;  
\begin{equation}\label{ENERGYEQ} 
d\rho/d\tau = -3aH(p+\rho) + a^{-1}V_{f}(d\phi/d\tau)^{2}
\end{equation}
Eqs. (\ref{SCALAREQ},\ref{HUBBLE},\ref{ENERGYEQ}) are sufficient 
to solve the model; we note that as $\phi$ decreases kinetic energy 
is taken from the scalar field and given to the fluid. The Einstein   
equations remain valid.

For calculation it is convenient to use 
\begin{equation} 
x \equiv \ln a
\end{equation}
as evolution parameter instead 
of $\tau$ and we use the notation $' \equiv d/dx$. The equations of 
motion then become  
\begin{equation}\label{SCALAREQ2} 
\phi''+ \frac{H'}{H}\phi'+ 3\phi' + V_{1}/H^{2}+ V_{f}\phi'/H = 0,
\end{equation}
\begin{equation}\label{HUBBLE2} 
H^{2}= \kappa (V + \rho)/(3-\frac{\kappa}{2}\phi'^{2}), 
\end{equation}
\begin{equation}\label{ENERGYEQ2} 
\rho' = -4\rho + HV_{f}\phi'^{2},
\end{equation}
where in the last equation we have used $p = \frac{1}{3}\rho$ for the  
relativistic fluid. In these equations $V,V_{1},V_{f}$ are given 
functions of $\phi$ only, and the solution of the equations gives 
$\rho(x),\phi(x),H(x)$ as functions of $x = \ln a$ only. We can find 
the value of $\tau$ as a function of $x$ and thus of $a$ by
\begin{equation}\label{TAUx} 
\tau = \int^{x} \frac{dx}{H} \exp(-x). 
\end{equation}

We note that the equations scale in the following sense: if we have 
a solution with a certain $V_{0}$, Eq.(\ref{POTENL2}), with a certain 
$\phi(x), \rho(x), H(x)$ and $c$ is any constant {\it then} 
$c^{2}V_{0}, \phi, c^{2}\rho, cH$ is also a solution. This is  
convenient because if we have a solution with a particular $V_{0}$ 
then, by adjustment of $c$, we can get a solution appropriate to the 
era we wish to consider.

\subsubsection{$\phi > \phi_{A}$; power-law inflation}
\label{REHEATA1}
In this region $\rho = V_{f} =0, V = U\exp(\lambda\phi),  
V_{1} = U\lambda\exp(\lambda\phi)$ and 
Eqs.(\ref{SCALAREQ2},\ref{HUBBLE2}) yield
\begin{equation}\label{SCALAREQ3} 
\phi''/(1-\frac{\kappa}{6}\phi'^{2}) + 3(\phi'+\lambda/\kappa) = 0.
\end{equation}
The simplest solution is
\begin{equation}\label{SOLN1} 
\phi' = -\frac{\lambda}{\kappa}, 
H^{2}=2U\exp(\lambda\phi)/(6-\frac{\lambda^{2}}{\kappa}).
\end{equation}
Letting $\Lambda \equiv \lambda^{2}/\kappa$ it follows that 
$H \propto \exp(-\frac{\Lambda x}{2}), \tau -\tau_{p} \propto  
\exp[(\frac{\Lambda}{2}-1)x],a \propto (\tau -\tau_{p})^{p}, 
p \equiv 2/(\Lambda-2), \tau_{p} = constant$. With $\Lambda < 2$ this 
is the well known power-law inflation due to an exponential 
potential.
\subsubsection{$\phi < \phi_{A}$}
In this region we need to use numerical compution to solve even the  
unperturbed equations whose solution we shall denote by $\phi_{0},  
\rho_{0}$. In our numerical computation the unperturbed solution 
and the perturbation to it are evolved simultaneously, and we proceed 
now to the formalism for this latter.
\subsection{The perturbed equations}\label{REHEATB}
Perturbations to the energy-momentum tensor density are coupled to   
scalar perturbations of the metric tensor. For these we work in a 
synchronous coordinate system and follow in the main the formalism 
of Grishchuk \cite{GRI}. The metric is  
\begin{equation}\label{METRIC} 
ds^{2} = -a^{2}d\tau^{2} + a^{2}[\delta_{ij} + h_{ij}(x)]dx^{i}dx^{j}.
\end{equation}
$h_{ij}(x)$ are the relevant perturbations which can be expressed as                                        
a linear superposition for the various wave numbers {\bf k} of 
\begin{equation}\label{Hij} 
h_{ij}({\bf k},x) = h({\bf k},\tau)Q\delta_{ij} + 
                    h_{l}({\bf k},\tau)k^{-2}Q_{,i,j},
\end{equation}
where $Q$ is a superposition of the spatial wave function solutions 
$\exp(\pm i{\bf k.x})$ so that $Q_{,i,j}=-k_{i}k_{j}Q$. The total 
perturbation can be found by solving for $h,h_{l}$ and this we shall  
do. We write the {\bf k}-components of the {\it perturbation} to the 
energy-momentum tensor as  
\begin{equation}\label{Tpert1} 
T_{k0}^{0}=\bar \rho_{1}Q, T_{ki}^{0} = -T_{k0}^{i} = 
a^{-2}\alpha\bar \xi'Q_{,i},
\end{equation}
\begin{equation}\label{Tpert2} 
T_{ki}^{j} = (\bar p_{1}+\bar p_{l})Q\delta_{i}^{j} + 
k^{-2}\bar p_{l}Q_{i}^{j}.
\end{equation}
where $\alpha \equiv dx/d\tau = \frac{da}{d\tau}/a = aH$, and where 
$\bar \rho_{1}$ and $\bar p_{1},\bar p_{l}$ are density and pressure 
perturbation components appropriate to {\bf k}. Then the 
perturbed Einstein equations where the variables are functions of 
$x=\ln a$ are 
\begin{equation}\label{pertEINST1} 
3h' + (k/\alpha)^{2}h - h_{l}' = 
\kappa(a/\alpha)^{2}\bar \rho_{1},
\end{equation}
\begin{equation}\label{pertEINST2} 
h' = \kappa\bar \xi'
\end{equation}
\begin{equation}\label{pertEINST3} 
-h'' - \frac{H'}{H}h' - 3h'= 
\kappa(a/\alpha)^{2}\bar p_{1},
\end{equation}
\begin{equation}\label{pertEINST4} 
\frac{1}{2}\lbrack h_{l}'' + \frac{H'}{H}h_{l}' + 3h_{l}' - 
(k/\alpha)^{2}h\rbrack = \kappa(a/\alpha)^{2}\bar p_{l}.
\end{equation}
In what follows we shall denote the {\bf k}-component perturbation to 
the scalar field by $\phi_{1}$ and the {\bf k}-component fluid 
perturbations by $\rho_{1},p_{1},\xi'$ and $p_{l}$. For the perfect 
fluid, which we assume, $p_{l} = 0$  and the corresponding quantity 
vanishes for the scalar field; so  $\bar p_{l} = 0$. The right hand 
sides of Eqs.(\ref{pertEINST1}-\ref{pertEINST4}) are given by: 
\begin{equation}\label{rhsEINST1} 
(a/\alpha)^{2}\bar \rho_{1} = \phi_{0}'\phi_{1}' + 
(\phi_{1}V_{1}+\rho_{1})/H^{2},
\end{equation}
\begin{equation}\label{rhsEINST2} 
\bar \xi' = \xi' - \phi_{0}'\phi_{1},
\end{equation}
\begin{equation}\label{rhsEINST3} 
(a/\alpha)^{2}\bar p_{1} = \phi_{0}'\phi_{1}' -
(\phi_{1}V_{1}-\frac{1}{3}\rho_{1})/H^{2},
\end{equation}
\begin{equation}\label{rhsEINST4} 
(a/\alpha)^{2}\bar p_{l} = 0.
\end{equation}
Inserting these into the preceeding equations we have a set of   
differential equations capable of solution when we have added the 
following perturbed scalar field equation;
\widetext
\begin{equation}\label{pertSCALAR} 
\phi_{1}''+ \frac{H'}{H}\phi_{1}'+ 3\phi_{1}' + 
\frac{\phi_{1}}{H^{2}}\frac{d^{2}V}{d\phi^{2}} +  
\lbrack\frac{k}{aH}\rbrack^{2}\phi_{1} + 
\frac{1}{2}(3h'-h_{l}')\phi_{0}' = 
-\lbrack\frac{dV_{f}}{d\phi}\phi_{0}'\phi_{1}+V_{f}\phi_{1}'\rbrack /H
\end{equation}
This equation combined with the (perturbed) Bianchi identities 
yields the perturbed fluid energy equation as: 
\begin{equation}\label{pertENERGY} 
\rho_{1}' + 4\rho_{1} - (k/a)^{2}\xi' + 
\frac{2}{3}\rho_{0}(3h'-h_{l}') = 
\lbrack\frac{dV_{f}}{d\phi}\phi_{0}'^{2}\phi_{1}+
2V_{f}\phi_{0}'\phi_{1}'\rbrack H
\end{equation}

\narrowtext

As in the unperturbed case the physical assumption lies in the 
addition of the friction term to the scalar equation; the Einstein  
equations are preserved leading to the perturbed fluid energy  
equation, including the appropriate friction term which balances the 
one in the scalar equation.

\section{Gradual and sudden transitions}\label{TRANSITIONS}
\subsection{Inflationary era}\label{TRANSA}

The initial conditions for solving the equations in the transition 
region, $\phi_{A} > \phi > 0$, are given by the solution at the end, 
$\phi = \phi_{A}$, of the power law inflation in the region 
$\phi > \phi_{A}$. In that region we already have the unperturbed 
solution in \ref{REHEATA1} and shall now use 
well-known analytic methods to get the perturbed solution.

We shall use dotted quantities to denote differentiation with respect 
to $\tau$; for example
\begin{equation}
\dot h \equiv dh/d\tau
\end{equation}
From Eqs.(\ref{pertEINST2},\ref{rhsEINST2})
\begin{equation}\label{EQ.phi}
\kappa\phi_{1} = -\dot h/\dot \phi_{0}
\end{equation}
It can be shown that for the scalar field only case \cite{GRI}
\begin{equation}
\ddot \mu + 
\mu\lbrack k^{2}-\ddot{(a\sqrt{\gamma})}/(a\sqrt{\gamma})\rbrack=0
\label{eq.ddotmu}
\end{equation}
\begin{equation}
\mu  \equiv  \frac{a}{\alpha\sqrt{\gamma}}(\dot h + \alpha\gamma h),
\label{EQ.mu}
\end{equation}
\begin{equation}
\gamma  \equiv  1 - \dot \alpha/\alpha^{2},
\label{eq.gamma}
\end{equation}
where, as defined in \ref{REHEATB}, $\alpha = \dot a/a$. 
The scale factor is given by Eq.(\ref{aPOWLAW})
so that $\gamma = (1 + p)/p = {\rm constant}$.

Integration of Eq.(\ref{EQ.mu}) gives $h$ as
\begin{equation}\label{EQ.h}
h = \frac{\alpha}{a} \lbrace \sqrt{\gamma} \int^{\tau}
\mu(k,\tau)d\tau + C_{i} \rbrace ,
\end{equation}
where $C_{i}$ is an integration constant subsuming, when $\gamma$ is  
constant, the lower limit of the integration. 
$h = \frac{\alpha}{a}C_{i}$ is a solution of $\mu = 0$ which can be  
eliminated by a coordinate transformation allowed by the residual 
gauge freedom of the synchronous gauge \cite{GRI}; it gives 
zero contribution to gauge invariant variables \cite{MUK,HEN}. 
We drop this term here but we shall revisit it later in conjunction 
with the results.  

All the above development was non-quantum mechanical. We shall now  
briefly remark on the quantum basis of the phenomena. We denote the 
corresponding quantum field theory quantities by a tilde:
\begin{equation}\label{eq.htil}
\tilde{h} = \frac{\alpha}{a} \sqrt{\gamma} \int^{\tau} 
\tilde{\mu}(k,\tau)d\tau   
\end{equation}
\begin{equation}\label{EQ.mutil}  \tilde{\mu} = 
N\int \frac{d^{3}k}{(2\pi)^{\frac{3}{2}}\sqrt{2k}} 
\lbrack c_{k}\mu_{1}(k,\tau)\exp(i{\bf k.x}) 
+ h.c. \rbrack
\end{equation}
where $c_{k}$ is a quantum annihilation operator, $\lbrack
c_{k},c_{k'}^{\dag}\rbrack=\delta^{3}({\bf k-k'})$,
and $\mu_{1}(y), y = |k(\tau_{p} - \tau)|$, is that solution of the
Bessel equation (\ref{eq.ddotmu}) such that
\begin{equation}\label{eq.limits}
\mu_{1}(y) \to e^{-iy}, y \to \infty 
\end{equation}
thus corresponding for large $k$ to the usual mode function of 
quantum mechanical plane waves.

$N$ is a normalization factor whose determination gives the 
absolute magnitude of the observed density perturbations (in terms 
of the model parameters) given the assumption that these come from 
a primordial vacuum with zero quantum occupation number. N has been 
determined as \cite{GRI,HEN}
\begin{equation}\label{eq.N}
N = \sqrt{2\kappa} = \sqrt{16\pi G}.
\end{equation} 

For this paper, which is mainly about sudden transition versus 
gradual transition amplitudes, we do not need to discuss the quantum 
aspects further. 

We can now proceed with the evaluation of the metric scalar 
components at the end of inflation. The exact expression for 
the solution $\mu_{1}$ is, with $n = \frac{1}{2} - p$, 
\begin{equation}\label{EQ.mu1}
\mu_{1}(y) = \sqrt{\frac{\pi y}{2}}(J_{n} - iY_{n})
\exp\lbrack-i(\frac{1}{2}n\pi + \frac{1}{4}\pi)\rbrack
\end{equation}    
Exact power-law inflation ends and the transition region begins at 
$\phi=\phi_{A}$; other variables at that surface,$A$ , we denote by 
the suffix $A$. For example: $y_{A} = |k(\tau_{p} - \tau_{A})|$. For 
small enough values of $y_{A}$ we can expand $\mu$ given by 
Eq.(\ref{EQ.mu1}) in an ascending power series and just consider the 
first one or two terms. Our first interest is in 
values of $k$ of importance in COBE and other observations. 
These are of the order of magnitude (the relevant 
meaning here is being within a few factors of ten) of
\begin{equation}\label{EQ.k1} 
k_{1} \equiv a'_{1}/a_{1} = a_{1}H_{1}
\end{equation}
where $a_{1},H_{1}$ denote the values of the scale factor and Hubble
at the time, $\tau_{1}$, when the matter era begins. Thus for such
values of $k$ the values of $y_{A}$ are of order $10^{-n} , n > 10$. 
Then we expand $\mu$ in a power series with leading terms
\begin{equation}
\mu_{1}(y) = M(p)\lbrack y^{p}-\frac{y^{p+2}}{2(2p+1)}.....\rbrack,
\label{eq.mu1y}
\end{equation}
\begin{equation}
M(p) \equiv -2^{-p}\Gamma(\frac{1}{2}-p)\exp(-i(1-p)/2) /\sqrt{\pi}.
\label{eq.Mp}
\end{equation}
The corresponding expansion of $h$ from Eq.(\ref{EQ.h}), with
 $C_{i}=0$, is 
\begin{equation}\label{EQ.hy} h = (\sqrt{\gamma}a)^{-1}
M(p)\lbrack y^{p} - \frac{(p+1)y^{p+2}}{2(2p+1)(p+3)}.....\rbrack
\end{equation}
We need these expressions at $y=y_{A}$ to find the initial values for  
the numerical solution of the coupled differential equations in
$h,h_{l},\phi_{1},\rho_{1}$. Noting that the first term in the 
expansion of $h$ 
\begin{equation}\label{SIGMA} 
\Sigma \equiv (\sqrt{\gamma}a)^{-1}M(p)y^{p} 
\end{equation}
is a constant, we find to first order in $y_{A}^{2}$ the initial   
values
\begin{equation}\label{INIT1} 
h=\Sigma + y_{A}^{2}P_{1}\Sigma 
\end{equation}
\begin{equation}\label{INIT2} 
h'=\frac{2}{p}y_{A}^{2}P_{1}\Sigma 
\end{equation}
\begin{equation}\label{INIT3} 
h_{l}'=\frac{1}{p(2p+1)}y_{A}^{2}\Sigma  
\end{equation}
\begin{equation}\label{INIT4} 
\phi_{1}=\frac{2}{p\lambda}y_{A}^{2}P_{1}\Sigma  
\end{equation}
\begin{equation}\label{INIT5} 
\phi_{1}'=\frac{2}{p}\phi_{1}
\end{equation}
\begin{equation}\label{INIT6} 
\rho_{1}'=0
\end{equation}
where
\begin{equation}\label{INIT7} 
P_{1}=-(p+1)/[2(p+3)(2p+1)]
\end{equation}

\subsection{Reheating and the radiation era}\label{TRANSB}

In Eqs.(\ref{pertEINST1}-\ref{pertEINST4}) $h$ occurs only when 
multiplied by $(k/\alpha)^{2}\equiv y^{2}/p^{2}$. This and the initial  
conditions above lead us to rewrite the equations in terms of 
renormalized variables which are of order zero in $y^{2}$. These are: 

$\hat h=(\alpha_{A}/k)^{2}(h-\Sigma)/\Sigma$
, $\hat h_{l}'=(\alpha/k)^{2}h_{l}'/\Sigma$
, $\hat\phi_{1}=(\alpha_{A}/k)^{2}\phi/\Sigma$ 
, $\hat\rho_{1}=(\alpha_{A}/k)^{2}\rho/\Sigma$
, $\hat\xi_{1}=(\alpha_{A}/k)^{2}\xi/\Sigma$.

Except for $h_{l}'$ the renormalization is by the same constant factor 
coinciding with the factor for $h_{l}'$ at $A$. The initial conditions 
for the hatted variables are given by the replacement 
$y_{A}^{2}\Sigma \to p^{2}$ on the right hand sides of 
Eqs.(\ref{INIT1}-\ref{INIT6}). The hatted versions of the last two 
Einstein equations are 
\begin{equation}\label{EQ3} 
\hat h''+(\frac{H'}{H}+ 3)\hat h'= 
-\kappa[\phi_{0}'\hat \phi_{1}'-
(\hat \phi_{1}V_{1}-\frac{1}{3}\hat \rho_{1})/H^{2}],
\end{equation}
\begin{equation}\label{EQ4} 
(\hat h_{l}')'+(1-\frac{H'}{H})\hat h_{l}'-1=0 
\end{equation}
We shall supplement Eqs.(\ref{EQ3},\ref{EQ4}) by 
Eqs.(\ref{EQ5},\ref{EQ6}) below, which are the hatted versions of  
Eqs.(\ref{pertSCALAR},\ref{pertENERGY}). We shall omit the explicit 
$k^{2}$ terms because these are of order 
$(k/\alpha)^{2}\equiv y^{2}/p^{2}$ smaller than the other terms in 
the equation for both cases. This is a consistent, controlled 
approximation, extremely good for COBE and related observations. 
\begin{equation}\label{EQ5} 
\hat \phi_{1}''+(\frac{H'}{H}+ 3)\hat\phi_{1}' + 
\frac{\hat\phi_{1}}{H^{2}}\frac{d^{2}V}{d\phi^{2}} +  
\frac{1}{2}[3\hat h'-((aH)_{A}/(aH))^{2}\hat h_{l}']\phi_{0}' = 
-\lbrack\frac{dV_{f}}{d\phi}\phi_{0}'\hat \phi_{1}+
V_{f}\hat \phi_{1}'\rbrack /H
\end{equation}  
\begin{equation}\label{EQ6} 
\hat\rho_{1}' + 4\hat\rho_{1} + 
\frac{2}{3}\rho_{0}[3\hat h'-((aH)_{A}/(aH))^{2}\hat h_{l}'] = 
\lbrack\frac{dV_{f}}{d\phi}\phi_{0}'^{2}\hat\phi_{1}+
2V_{f}\phi_{0}'\hat\phi_{1}'\rbrack H
\end{equation}
\narrowtext
Eqs.(\ref{EQ3})-(\ref{EQ6}), with the initial conditions specified 
above, have to be solved for the perturbations. It is necessary to 
know the unperturbed solutions  $\phi_{0}$, $\rho_{0}$ and $H$ which 
enter the equations; these are given by eqs.(\ref{SCALAREQ2})- 
(\ref{ENERGYEQ2}) and their initial conditions at the beginning of 
reheating are given in \ref{REHEATA1}. We evolve all the variables,  
both the perturbed and the unperturbed, together as a set of coupled 
differential equations with evolution parameter $x=\ln a$. The end 
of the transition period and the beginning of the radiation era are 
signalled by $\phi_{0}$ arriving at a negligable value; $\phi_{1}$ 
accompanies $\phi_{0}$ into oblivion. 

Our objective is to compare the radiation era amplitudes resulting  
from this gradual transition to those resulting from a sudden 
transition.

\subsubsection{Sudden transition to the radiation era}
\label{TRANSB1}

For this we use the matching conditions given by Deruelle and 
Mukhanov \cite{DER}; conditions (their eqs. 4.5a,b) in the  
longitudinal gauge can equivalently be given in terms of the gauge 
invariant variable $\Phi$. They are that $\Phi$ and $\Gamma$ are  
continuous across the matching surface $\tau=\tau_{2}$:   
\begin{equation}\label{CONT1} 
\Phi_{+}\equiv \Phi(\tau_{2+})=\Phi(\tau_{2-})\equiv \Phi_{-}  
\end{equation}
\begin{equation}\label{CONT2} 
\Gamma_{+}\equiv \Gamma(\tau_{2+})=\Gamma(\tau_{2-})\equiv\Gamma_{-}  
\end{equation}
where
\begin{equation}\label{GAMMA} 
\Gamma \equiv (\Phi' + \Phi + \epsilon^{2}\Phi)/\gamma, 
\end{equation}
\begin{equation}\label{EPSILON} 
\epsilon = k/(\alpha\sqrt{3}). 
\end{equation}
Additionally, more obviously, $a$ and $\alpha=\dot a/a=aH$ must also be 
continuous across the surface.

Density perturbations can be described by the mode function $\nu$,  
corresponding to the $\mu_{1}$ of the inflation era, such that  
\cite{GRI,HEN}
\begin{equation}\label{NU1} 
\ddot\nu + \frac{1}{3}k^{2}\nu=0. 
\end{equation}
In terms of the synchronous gauge variables
\begin{equation}\label{NU2} 
\nu\equiv\frac{a}{\alpha}(\dot h+\alpha\gamma h)\equiv a(h'+\gamma h) 
\end{equation}
and the gauge invariant variable $\Phi$ is given by \cite{MUK}
\begin{equation}\label{PHIh} 
\Phi = h-(\alpha/k)^{2}h_{l}' 
\end{equation}
so that from Eqs.(\ref{pertEINST1}-\ref{pertEINST4})
\begin{equation}\label{PHINU} 
\Phi = (\nu-\nu')/a\epsilon^{2} 
\end{equation}
Non-physical synchronous gauge modes (that is of the form 
$h=\alpha/a\times constant$) give zero contribution to both $\nu$ and 
$\Phi$. We write the solution of Eq.(\ref{NU1}) as 
\begin{equation}\label{NU3} 
\nu = C\cos (k(\tau - \tau_{2})/\sqrt{3}) - 
S\sin (k(\tau - \tau_{2})/\sqrt{3}).
\end{equation}
so that 
\begin{equation}\label{PHI+} 
\Phi_{+} = (\epsilon_{2}S + C)/(a_{2}\epsilon_{2}^{2}),
\end{equation}
\begin{equation}
\Gamma_{+} = -\lbrack C(1-\epsilon_{2}^{2}) + \epsilon_{2} 
S(1-\frac{1}{2}\epsilon_{2}^{2})\rbrack /(a_{2}\epsilon_{2}^{2}).
\label{GAMMA+}
\end{equation}
From the continuity of $\Phi$ and $\Gamma$ across the interface these  
expressions enable the determination of $C,S$ and thus the density 
perturbation in the radiation era. So we need the values of  
$\Phi_{-}$ and $\Gamma_{-}$ from the inflationary era solutions.
From Eqs.(\ref{INIT1}, \ref{INIT3},\ref{PHIh}), interpreted now as being 
at the surface $\tau=\tau_{2}$, we find that to lowest order in 
$(k/\alpha)^{2}$
\begin{equation} 
\Phi_{-} =\frac{p+1}{2p+1} \Sigma;\Gamma_{-}= \frac{p}{2p+1} \Sigma.
\end{equation}
Solving Eqs.(\ref{CONT1},\ref{CONT2}) straightforwardly gives to 
lowest order
\begin{equation}\label{CS1} 
C=-\epsilon_{2}S=2a_{2}\Sigma.
\end{equation}
For $k$ relevant to CMBR fluctuation observations $\epsilon_{2}$ is 
of order $10^{-n}$ where $n>10$ and thus away from the interface the 
term in $S$ dominates the radiation era density perturbations.

\subsubsection{Gradual reheating transition}\label{TRANSB2}

For our reheating model, or any such continuous transition, no 
question of sophisticated implementation of Lichnerowicz conditions 
at an interface arises. There is no interface; the physical variables 
arising from the solutions of the equations of motion are all   
continuous through space-time.

We shall consider the case where the transition ends at the same 
surface, $\tau=\tau_{2}$, as that of \ref{TRANSB1}; $\phi$ is 
negligable for $\tau\geq\tau_{2}$ and the system is in the radiation 
era. The transition begins at the end of pure power law inflation, 
$\tau=\tau_{A}$. We note here that $\Sigma$ of Eq.(\ref{SIGMA}) is a  
constant of the power-law inflation era. Below we shall refer to a 
$\Sigma$ evaluated at $\tau_{A}$ at the end of power-law inflation
at the beginning of the gradual transition; while in  
 \ref{TRANSB1} we referred to $\Sigma$ evaluated at that appropriate   
(sudden) end of power-law inflation, $\tau_{2}$; these have the   
same name and are indeed the same numbers.

The radiation era analysis is of course precisely the same as in 
\ref{TRANSB1}, with Eqs.(\ref{PHIh},\ref{PHINU}) giving  
\begin{equation}\label{CS4} 
3\Phi+\Phi' =-\frac{3}{ak^{2}}\ddot\nu = \frac{1}{a}\nu ;
\end{equation}
so at the end of the transition, $\tau=\tau_{2}$, which is the 
beginning of the radiation era
\begin{equation}\label{CS5} 
C/a_{2} = 3\Phi_{2}+\Phi_{2}',
\end{equation}
\begin{equation}\label{CS6} 
(C+\epsilon_{2}S)/(a_{2}\epsilon_{2}^{2}) = \Phi_{2}. 
\end{equation}
Putting into the RHS of the last two equations the computed values 
at the end of the transition we can find $C$ and $S$. In fact, 
because of the very small magnitude of $\epsilon_{2}$, the last 
equation implies that to lowest order in $\epsilon_{2}$ 
\begin{equation}\label{CS7} 
\epsilon_{2}S = -C 
\end{equation}
since otherwise $\Phi_{2}$ would take a large value unattainable in 
the transition. The result we have arrived at is 
that for the continuous case the density perturbation in the 
radiation era is determined by the value of 
$3\Phi_{2}+\Phi_{2}'$ at the end of the gradual transition. 

By numerical integration of Eqs.(\ref{EQ3}-\ref{EQ6}) with the stated 
initial conditions we determine $\Phi_{2}$ and $\Phi_{2}'$ in terms 
of $\Sigma$ using Eq.(\ref{PHIh}). We can then compare the results 
for $S$ and $C$, determining the radiation era perturbations, from 
the sudden and gradual transitions.

\subsubsection{Comparison of sudden with gradual transition}
\label{TRANSB3}

For the sudden transition we have, Eq.(\ref{CS1}),
\begin{equation}\label{CS8} 
-\epsilon_{2}S/a_{2}\Sigma =C/a_{2}\Sigma = 2
\end{equation}

For the gradual transition we denote 
\begin{equation}\label{result1} 
result \equiv (3\Phi_{2}+\Phi_{2}')/\Sigma
\end{equation}
so that
\begin{equation}\label{result2} 
-\epsilon_{2}S/a_{2}\Sigma =C/a_{2}\Sigma =result
\end{equation}

Thus if $result = 2$ the two transitions agree.

We have calculated the gradual transition for two values of the 
inflation power-law exponent, $p$, each for a number of values of  
the friction coefficient, $f$, with the results shown in the Tables.
It may be noted that for the larger values of $f$ ,($f > 1$) quoted 
the system grinds to a halt at the bottom of the potential well  
with no significant oscillations; this of course does not resemble  
any usually envisaged reheat mechanisms \cite{KOF}. For smaller $f$ 
oscillations occur and it is found, as we would expect, that  
their number increases rapidly as $f$ decreases. Our numerical 
computation methods fail for $f$ significantly smaller than the last 
value in the Tables, $f = .01$.

As stated previously the transition is deemed to have come to an end 
when the value of $\phi_{0}$ is sufficiently small. There is not a 
precise criterion and we display results for various values of 
$\phi_{0}(end)$. We see that always $result=2$ to very high 
accuracy, in agreement with the sudden transition.

This establishes the agreement (for appropriate wavelengths) in the  
following sense: whether there is a sudden transition at red-shift 
$z_{2}$ or our type of gradual transition ending at $z_{2}$, the 
resulting radiation era scalar perturbation is the same. 

\subsubsection{Synchronous gauge mode}
\label{TRANSB4}

As set out in Eqs.(\ref{EQ.phi})-(\ref{EQ.h}) there is in the 
inflationary era (as in other eras) a 'synchronous gauge mode' given 
by
\begin{equation}\label{hSYNCH} 
h=\frac{\alpha}{a}C_{i}
\end{equation}
where $C_{i}$ is a constant. This mode is non-physical as it 
contributes zero to the gauge invariant variable $\Phi$, given by 
Eq.(\ref{PHIh}), as well as to other gauge invariant variables, and 
also it can be eliminated by a coordinate transformation making 
use of the residual gauge freedom of synchronous gauges.

Whole these remarks justify our ommission of this term, we have 
investigated further so as to provide a comment on the mechanism 
of energy transfer inthe gradual transition of our model. Our theory 
is a linear perturbation theory , so we can completely investigate 
the consequences of a synchronous gauge mode separately from the 
physical modes previously considered by inserting solely 
Eq.(\ref{hSYNCH}) and its derivative and 
\begin{equation}\label{hlSYNCH} 
h_{l}'=(k/\alpha)^{2}\frac{\alpha}{a}C_{i}
\end{equation}
as the initial conditions in Eqs.(\ref{EQ3})-(\ref{EQ6}). These 
correspond to zero $\Phi$ in the inflation era. For a perfect, that 
is coordinate independent, perturbed transition mechanism in 
Eqs.(\ref{EQ5}) and (\ref{EQ6}) the resulting $\Phi$ in the radiation 
era would be zero. To give an idea of the magnitudes involved we can 
put, formally, $(\frac{\alpha}{a})_{A}C_{i} = \Sigma$ at the 
beginning of the transition - that is as if this quantity began the 
transition with the same magnitude as the physical mode previously 
considered. We illustrate by giving two results, corresponding to  
the first and last lines of Table I, for the value of $\Phi$ at the 
end of the transition: 
\begin{equation}\label{f4SYNCH} 
f = 4 \Rightarrow \Phi_{2} 
= \Sigma \times 10^{-3},
\end{equation}
\begin{equation}\label{fp1SYNCH} 
f = .01 \Rightarrow \Phi_{2} 
=4.36 \Sigma \times 10^{-6}.
\end{equation}
The results for the physical mode were $2\Sigma$ for whatever $f$. 
Eqs. (\ref{f4SYNCH}) and (\ref{f4SYNCH}) illustrate firstly that 
there is a coordinate dependence in the transition mechanism when 
applied to the non-physical, coordinate dependent, perturbative mode 
and secondly that it decreases with $f$.

It is probable that other transition mechanisms expressible as terms   
in the Lagrangian, such as some of those which have been invoked to 
investigate particle production in the reheat phase \cite{TRA},
would transmit only coordinate independent information into 
coordinate independent variables.However investigation of these 
would involve different formalisms and calculations. We would not 
expect any different answers for the physical modes.

\subsection{Gravitational waves}\label{TRANSC}

Tensor perturbations are very much simpler to treat than scalar 
perturbations. After their generation by quantum fluctuations their 
development is governed solely by the cosmic scale factor $a$. 
Corresponding to Eq.(\ref{EQ.mutil}) we have in the power-law 
inflationary era 
\widetext
\begin{equation}\label{HIJTIL} 
\tilde{h}_{ij} = \sqrt{16\pi G}\sum_{\lambda =1}^{2}\int 
\frac{d^{3}k}{(2\pi)^{\frac{3}{2}}\sqrt{2k}a(\tau)} 
\lbrack a_{\lambda k}\epsilon_{ij}^{\lambda}(k)
\mu_{1}(k,\tau)\exp(i{\bf k.x}) + h.c. \rbrack,
\end{equation}
\narrowtext
where $a_{\lambda k}$ is the annihilation operator for the graviton
with polarization $\lambda$ and wave number ${\bf k}$,
and the polarization tensor satisfies 
$\sum_{i,j}\epsilon_{ij}^{\lambda}(k) \epsilon_{ij}^{\lambda'}(k) = 
2\delta_{\lambda \lambda'}$. $\mu_{1}$ given by Eq.(\ref{eq.mu1y}) is a 
solution, appropriate to the quantum mechanics as in 
Eq.(\ref{eq.limits}), of the tensor mode equation 
 \begin{equation}\label{EQ.mugrav} 
\ddot\mu + \mu(k^{2} - \ddot a/a) = 0 .
\end{equation}
This holds for any cosmic era, whatever may be the dynamics
 responsible for the particular form of $a(\tau)$ and in the 
radiation era, beginning at $a=a_{2}$, which defines $\tau=\tau_{2}$,  
we can write its solution as 
\begin{equation}\label{MURADERA} 
\mu = G_{+}\cos(k(\tau - \tau_{2})) - G_{-}\sin(k(\tau - \tau_{2})),
\end{equation}   
where $G_{+}$ and $G_{-}$ are constants to be determined by 
continuity, as set out below.

\subsubsection{Sudden transition}\label{TRANSC1}                  

If we assume a sudden transition from inflation to the radiation era  
at $a=a_{2}$ the matching conditions that determine   
$G_{+}$ and $G_{-}$ are that $\mu$ and $\dot\mu$ (equivalently 
$\mu'$) must be continuous across the interface; the continuity of 
$a$ and $\alpha$ is also a condition but we already hold these 
enforced as in \ref{REHEATB} \cite{DER,ABB}. Up to the second term 
of the power series expansion $\mu$ is given by Eq.(\ref{eq.mu1y}) 
and $\mu'$ by  
\begin{equation}
\mu'_{1}(y) = M(p)\lbrack y^{p}-\frac{(p+2)y^{p+2}}{2p(2p+1)}\rbrack,
\label{eq.mu1'y}
\end{equation}
and the required continuity gives 
\begin{equation}
G_{+} = M(p)y_{2}^{p}\lbrack 1-\frac{y_{2}^{2}}{2(2p+1)}\rbrack,
\label{G+}
\end{equation}
\begin{equation}
G_{-} =p M(p)y_{2}^{p-1}\lbrack 1-\frac{(p+2)y_{2}^{2}}{2p(2p+1)}\rbrack.
\label{G-}
\end{equation}
Due to the extreme smallness of $y_{2}$ for waves of observational 
interest $G_{-}$ dominates and to lowest order 
\begin{equation}
G_{+} = 0,
\label{G+0}
\end{equation}
\begin{equation}
G_{-} =p M(p)y_{2}^{p-1}.
\label{G-0}
\end{equation}

\subsubsection{Comparison of gradual with sudden transition}
\label{TRANSC2}

At the end of power-law inflation, beginning the gradual transition
\begin{equation}
\mu(\tau_{A}) = 
M(p)y_{A}^{p}\lbrack 1-\frac{y_{A}^{2}}{2(2p+1)}\rbrack,
\end{equation}
\begin{equation}
\mu'(\tau_{A}) = 
M(p)y_{A}^{p}\lbrack 1-\frac{(p+2)y_{A}^{2}}{2p(2p+1)}\rbrack.
\end{equation}
Also we shall below use the fact (see Eq.(\ref{SIGMA})) that 
\begin{equation}\label{SIGMAG}
\Sigma_{G}\equiv = M(p)y^{p}/a = \sqrt{\gamma}\Sigma
\end{equation}
is a constant of the inflationary era.
It is convenient to consider the function $\lambda\equiv\mu/a$ so 
that from Eq.(\ref{EQ.mugrav})
\begin{equation}
\ddot\lambda+2\dot\lambda\dot a/a+k^{2}\lambda=0
\end{equation}
Generally the solution is not trivial, but working to lowest order, 
putting $k^{2}=0$, we get 
\begin{equation}
a^{2}\dot\lambda=constant
\end{equation}
and, again to lowest order, at the end of the power-law inflation
\begin{equation}
\lambda=M(p)y^{p}/a \Rightarrow \dot\lambda_{A}=0.
\end{equation}
Then $\lambda$ is constant throughout the transition, $\mu\propto a$,  
as it also is in the inflationary phase, though for different 
$a(\tau)$; in both cases $\lambda = \Sigma_{G}$.  
This implies that, to lowest order, where $\mu$ and 
$\mu_{1}$ are the gradual transition and inflationary mode functions 
respectively: 
\begin{equation}
\mu(a_{2})=\mu'(a_{2})=\mu_{1}(a_{2})=\mu'_{1}(a_{2})
\end{equation}
since $a'=a$.
Thus Eqs.(\ref{G+0},\ref{G-0}) hold as in the sudden transition case. 

This establishes the same result for tensor perturbations as was 
established for scalar perturbations, and in the sense stated at the 
end of section \ref{TRANSB3}. For scalar perturbations  
it was likewise justifiable to neglect terms of order 
$(k/\alpha)^{2}$ in certain equations for waves with $k$ relevant   
to current observations. We may note that this easy way of finding 
the result in the tensor case is not only due to the relative  
simplicity of the governing equation (\ref{EQ.mugrav}) but also to 
the assumption that the inflation is power-law. However we would  
expect it to hold more generally.

\section{Discussion}\label{DISCUSSION}

As the transition proceeds both the unperturbed, $\phi_{0}$, and the 
perturbed, $\phi_{1}$, scalar fields diminish so that at some 
sufficiently small value we can use the perfect fluid model of the 
radiation era. In the latter part of the radiation era, as the  
the matter era is approached, the amplitude of the 
density perturbation is specified by the single quantity 
$result \equiv 3\Phi + \Phi' = 3\Phi + (aH)^{-1}d\Phi /d\tau $,   
evaluated at the end of the transition. 

An extension of the length of the transition, 
specified in the Tables by the number of e-foldings, can always give 
values of $|\phi_{0}|$ as small as we please until the computation 
runs out of numerical accuracy. We have chosen the varying examples 
in the tables to demonstrate that the lack of precision in the 
$|\phi_{0}|$ criterion is not important in the sense that 
$result \equiv (3\Phi_{2}+\Phi_{2}')/\Sigma$  is very near to 2 for 
many different 'end' values of $|\phi_{0}|$. 

Though $\Phi_{2}/\Sigma$ and $\Phi_{2}'/\Sigma$ vary in the third 
decimal place the $result$ is much more constant, and we can 
understand this: In the radiation era by Eqs.(\ref{CS4},\ref{NU3})  
$(3\Phi+\Phi')/\Sigma = -\epsilon S/a\Sigma +O(\epsilon^{2})$ and 
the first term is constant since $a \propto \tau$ and 
$\epsilon \propto \tau$ and an equivalent result would apply towards 
the end of the transition region. Thus for the above calculations  
with small $k$ the value of $3\Phi+\Phi'$ is very constant 
over many e-foldings of the cosmic scale factor $a$
after the scalar field has decreased in magnitude by more than about 
4 powers of 10. This is a property of the early part of the 
radiation era and not a property of the particular transition 
mechanism adopted. 

Having concurred with some other authors \cite{DER,CAL,MS,GOTZ,LYTH}, 
through different considerations, that transition period detail has 
no influence on density perturbations important for the CMBR  
fluctations, we may ask for what wave numbers $k$ the detail is  
going to be important. The answer is likely to vary considerably  
with the assumed physics of the early universe. So we can only 
give an answer for power-law inflation followed by reheating. For  
illustration it is simplest to consider the case of gravitational 
waves as in \ref{TRANSC}. To get the gradual-sudden equivalence 
there required that $p^{2}k^{2}/\alpha^{2} \ll 1$ throughout the 
transition region. We can conclude that 
for $k/k_{1} \approx 10^{23} h^{-.5}$ the sudden result is still 
reasonably good; here $h$ is the Hubble parameter. $k_{1}$, given 
by Eq.(\ref{EQ.k1}, for redshift $z_{1}=10^{3.3}$, being the value  
around which we did our tabulated calculations above, 
specifies the region of wave numbers of 
importance in CMBR fluctuations. Translated into energy terms this 
means that gravitational waves with $k<10^{-11}h^{.5}eV$ would not be  
affected by reheating details. A calculation with density 
perturbations gives similar results.

\newpage
\begin{table}
\caption{$result \equiv (3\Phi_{2}+\Phi_{2}')/\Sigma$ at the 'end'
 (see text) of the gradual transition for values of the friction 
 coefficient, $f$; $R_{\phi}$ is the ratio of the unperturbed scalar 
 field value, $\phi_{0}$, at the transition 'end' to that at the 
 beginning; $x(end)$ gives the number of e-foldings of the 
transition. Conformal time inflation power $p=-1.113$.}
\label{T1} 
\begin{tabular}{|c|c|c|c|c|c|} 
$f$ & $x(end)$ & $|R_{\phi}|$ & $\Phi_{2}/\Sigma$ & 
$\Phi_{2}'/\Sigma$ & $|result-2|$\\ \hline
$4.0$ & $3.2$ & $0.29\times 10^{-10}$ & $0.66162$ & $0.01515$ & 
$<10^{-8}$\\ 
$2.0$ & $3.0$ & $0.20\times 10^{-23}$ & $0.66262$ & $0.01214$ & 
$<10^{-8}$\\  
$1.0$ & $3.0$ & $0..30\times 10^{-18}$ & $0.66343$ & $0.00971$ & 
$<10^{-8}$\\ 
$0.5$ & $3.0$ & $0.34\times 10^{-10}$ & $0.66337$ & $0.00990$ & 
$<10^{-8}$\\  
$0.1$ & $3.5$ & $0.18\times 10^{-6}$ & $0.66450$ & $0.00650$ & 
$1.1\times 10^{-8}$\\  
$0.05$ & $4.0$ & $0.14\times 10^{-6}$ & $0.66534$ & $0.00399$ & 
$<10^{-8}$\\  
$0.01$ & $5.0$ & $0.54\times 10^{-7}$ & $0.66540$ & $0.00380$ & 
$<10^{-8}$\\  
\end{tabular}
\end{table}
\begin{table}
\caption{$result \equiv (3\Phi_{2}+\Phi_{2}')/\Sigma$ at the 'end'
 (see text) of the gradual transition for values of the friction 
 coefficient, $f$; $R_{\phi}$ is the ratio of the unperturbed scalar 
 field value, $\phi_{0}$, at the transition 'end' to that at the 
 beginning; $x(end)$ gives the number of e-foldings of the 
transition. Conformal time inflation power $p=-1.05$.}
\label{T2} 
\begin{tabular}{|c|c|c|c|c|c|} 
$f$ & $x(end)$ & $|R_{\phi}|$ & $\Phi_{2}/\Sigma$ & 
$\Phi_{2}'/\Sigma$ & $|result-2|$\\ \hline
$4.0$ & $4.0$ & $0.61\times 10^{-10}$ & $0.66236$ & $0.01293$ & 
$<10^{-8}$\\ 
$2.0$ & $4.0$ & $0.21\times 10^{-17}$ & $0.66476$ & $0.00572$ & 
$<10^{-8}$\\  
$1.0$ & $3.0$ & $0.26\times 10^{-4}$ & $0.63595$ & $0.09215$ & 
$2.5\times 10^{-8}$\\ 
$0.5$ & $4.0$ & $0.11\times 10^{-15}$ & $0.66511$ & $0.00467$ & 
$<10^{-8}$\\  
$0.1$ & $4.5$ & $0.50\times 10^{-9}$ & $0.66566$ & $0.00303$ & 
$<10^{-8}$\\  
$0.05$ & $5.0$ & $0.48\times 10^{-10}$ & $0.66605$ & $0.00185$ & 
$<10^{-8}$\\  
$0.01$ & $6.0$ & $0.15\times 10^{-10}$ & $0.66608$ & $0.00176$ & 
$<10^{-8}$\\  
\end{tabular}
\end{table}

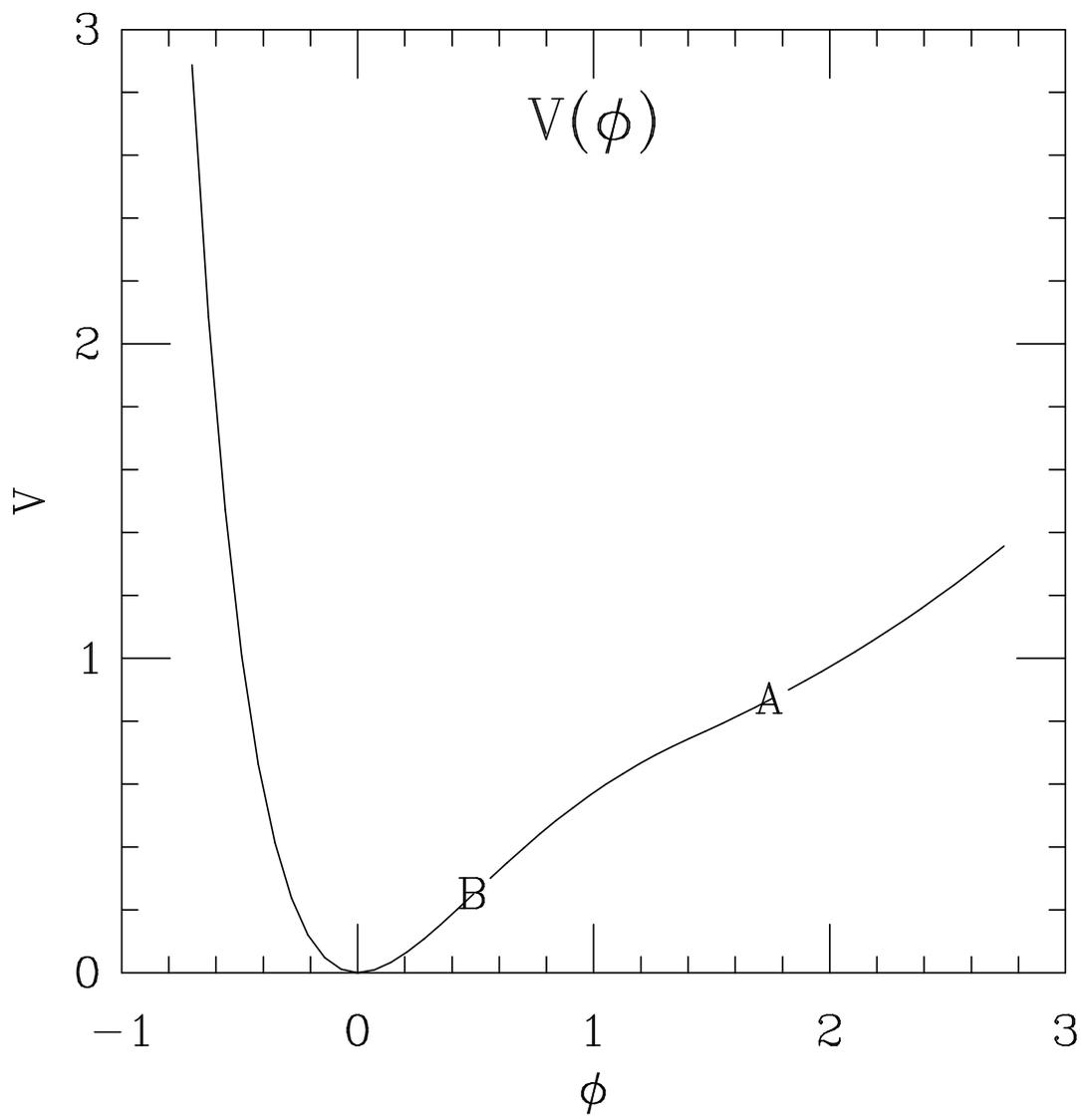
\end{document}